# Transformation media that turn a narrow slit into a large window

Xiaohe Zhang(张晓鹤), Huanyang Chen(陈焕阳), Xudong Luo(罗旭东)* and Hongru Ma(马红孺)

Department of Physics, Shanghai Jiao Tong University, Shanghai 200240, People's Republic of China

**Abstract**

Based on the transformation media theory, the authors propose a way to replace a wide window with a narrow slit filled with designed metamaterial to achieve the same transmittance as the one of the window. Numerical simulations for a two dimensional case are given to illustrate the ideas and the performance of the design.

*To whom correspondence should be addressed. E-mail: luoxd@sjtu.edu.cn

The concept of transformation media [1, 2] and its use in designing and realization of a microwave invisible cloak [3] has aroused great interest in the studies of controlling the waves. It has brought forth the possibility that people can design many kinds of devices with novel properties. Different designs of devices appeared in literature soon after the concept was initiated. The rotation cloaking was proposed which can rotate the appearance orientation of a cloaked object [4]. People also designed other kinds of transformation media, such as the wave concentrator [5], the wave shifter [6, 7, 8], the electromagnetic wormhole [9], etc. An acoustic transformation media concept [10] has also been proposed which can guide the acoustic waves freely. The cloak of mater waves was also given [11]. The concept that the transformed space is equivalent to a kind of special anisotropic material will produce more and more new kinds of transformation media that control the wave fields dramatically. In this paper, we suggest a kind of transformation media that can transmit the information outside a domain through a small slit, with the transmittance identical to the one of a big window.

We start from the general coordinate transformation with its Jacobian transformation matrix written as [12],

$$\Lambda_\alpha^{\alpha'} = \frac{\partial x^{\alpha'}}{\partial x^\alpha}. \qquad (1)$$

From the view of transformation optics, the transformed space is equivalent to a set of permittivity and permeability,

$$\varepsilon^{i'j'} = |\det(\Lambda_i^{i'})|^{-1} \Lambda_i^{i'} \Lambda_j^{j'} \varepsilon^{ij}, \quad \mu^{i'j'} = |\det(\Lambda_i^{i'})|^{-1} \Lambda_i^{i'} \Lambda_j^{j'} \mu^{ij}. \qquad (2)$$

The anisotropic material with the above tensors is "the transformation media", inside which the light will propagate with the same route as the one in the transformed space. From this general prescription, we now define the detailed coordinate transformation described in Fig. 1. Fig. 1(a) shows the original free space sandwiched between two obstacles to form a big window. Fig. 1(b) shows the transformed space compressed from the free space to form a narrow slit, which is also sandwiched between two obstacles. These kinds of sandwiching composition will separate the whole space into the upper space ($y, y' > \frac{d_1}{2}$) and the lower space ($y, y' < -\frac{d_1}{2}$). We suppose that the upper space is corresponding to an outside domain while the lower space is to an inside domain. The original rectangular region ($-\frac{w_1}{2} < x < \frac{w_1}{2}$ and $\frac{d_2}{2} < y < \frac{d_1}{2}$) is compressed into a trapezoidal region (see the region $\frac{d_2}{2} < y' < \frac{d_1}{2}$ in Fig. 1(b), for clarity, we define it as "Region I"). The corresponding mapping is,

$$x' = \frac{\frac{w_1 - w_2}{d_1 - d_2}(y - \frac{d_2}{2}) + \frac{w_2}{2}}{\frac{w_1}{2}} x, \quad y' = y, \quad z' = z. \qquad (3)$$

Another rectangular region ($-\frac{w_1}{2} < x < \frac{w_1}{2}$ and $-\frac{d_2}{2} < y < \frac{d_2}{2}$) is compressed into a smaller rectangular region ($-\frac{w_2}{2} < x' < \frac{w_2}{2}$ and $-\frac{d_2}{2} < y' < \frac{d_2}{2}$, defined as "Region II"). The corresponding mapping is,

$$x' = \frac{w_2}{w_1} x, \quad y' = y, \quad z' = z. \qquad (4)$$

A third rectangular region ($-\frac{w_1}{2} < x < \frac{w_1}{2}$ and $-\frac{d_1}{2} < y < -\frac{d_2}{2}$) is compressed into a trapezoidal region (see the region $-\frac{d_1}{2} < y < -\frac{d_2}{2}$ in Fig. 1(b), defined as "Region III"). The corresponding mapping is,

$$x' = \frac{-\frac{w_1 - w_2}{d_1 - d_2}(y + \frac{d_2}{2}) + \frac{w_2}{2}}{\frac{w_1}{2}} x, \quad y' = y, \quad z' = z. \qquad (5)$$

With the above coordinate transformation, the free space of the big window shown in Fig. 1(a) is transformed to the narrow slit shown in Fig. 1(b). According to the theory of transformation media, the slit can transmit the light just like the big window if the permittivity and permeability inside the slit is arranged with Eq. (2), which can be evaluated to give, for region I,

$$\vec{\mu} = \vec{\varepsilon} = \begin{pmatrix} \dfrac{\Delta_1^2 + \Gamma_1^2}{\Delta_1} & \dfrac{\Gamma_1}{\Delta_1} & 0 \\ \dfrac{\Gamma_1}{\Delta_1} & \dfrac{1}{\Delta_1} & 0 \\ 0 & 0 & \dfrac{1}{\Delta_1} \end{pmatrix}, \qquad (6a)$$

for region II,

$$\vec{\mu} = \vec{\varepsilon} = \begin{pmatrix} \dfrac{w_2}{w_1} & 0 & 0 \\ 0 & \dfrac{w_1}{w_2} & 0 \\ 0 & 0 & \dfrac{w_1}{w_2} \end{pmatrix}, \qquad (6b)$$

and for region III,

$$\vec{\mu} = \vec{\varepsilon} = \begin{pmatrix} \dfrac{\Delta_2^2 + \Gamma_2^2}{\Delta_2} & \dfrac{\Gamma_2}{\Delta_2} & 0 \\ \dfrac{\Gamma_2}{\Delta_2} & \dfrac{1}{\Delta_2} & 0 \\ 0 & 0 & \dfrac{1}{\Delta_2} \end{pmatrix}, \qquad (6c)$$

where the parameters are,

$$\Delta_1 = \frac{\dfrac{w_1-w_2}{d_1-d_2}(y'-\dfrac{d_2}{2})+\dfrac{w_2}{2}}{\dfrac{w_1}{2}}, \quad \Delta_2 = \frac{-\dfrac{w_1-w_2}{d_1-d_2}(y'+\dfrac{d_2}{2})+\dfrac{w_2}{2}}{\dfrac{w_1}{2}}, \qquad (7a)$$

$$\Gamma_1 = \frac{\dfrac{w_1-w_2}{d_1-d_2}}{\dfrac{w_1-w_2}{d_1-d_2}(y'-\dfrac{d_2}{2})+\dfrac{w_2}{2}} x' = \frac{\dfrac{w_1-w_2}{d_1-d_2}}{\Delta_1 \dfrac{w_1}{2}} x',$$

$$\Gamma_2 = \frac{-\dfrac{w_1-w_2}{d_1-d_2}}{-\dfrac{w_1-w_2}{d_1-d_2}(y'+\dfrac{d_2}{2})+\dfrac{w_2}{2}}x' = \frac{-\dfrac{w_1-w_2}{d_1-d_2}}{\Delta_2\dfrac{w_1}{2}}x'. \qquad (7b)$$

To demonstrate the useful properties of such a device, we will simply consider the two dimensional (2D) transverse electric (TE) polarization incident waves (whose electric field is along z-direction). The obstacles are mimicked using perfect electric conductor (PEC) boundaries. The following numerical results are obtained from the commercial finite-element solver COMSOL MULTIPHYSICS.

Before moving to the discussions of the properties of the device, we consider here the feasibility of such a device. To give a concrete example, we set $w_1 = 2\,m$, $d_1 = 0.5\,m$, $w_2 = 0.25\,m$ and $d_1 = 0.25\,m$. Here the unit of length is meter, which is typical for microwaves in this range. However, we can use smaller or bigger length as desired to provide the working wavelength to be in the same order of the slit width. Fig. 2 shows the spatial distribution of the components of the relative permeability and permittivity tensors. Fig. 2(a) shows the spatial distribution of $\mu_{xx}$, Fig. 2(b) shows the spatial distribution of $\mu_{yy} = \varepsilon_z$, and Fig. 2(c) shows the spatial distribution of $\mu_{xy}$. Except that there are some big values near the edges of the device, the anisotropic and inhomogeneous metamaterial is still feasible. The big values are due to the great anisotropy of the device. One can do some similar reduction as was done in the reduced cloak [3] to obtain a more feasible design. In addition, a design of optical conformal mapping [2] will also help to reduce the anisotropy of the metamaterial.

Now we come to the demonstration of the utility of this device. First, the electric line source in the free space is plotted in Fig. 3(a). Here the source carries a current of 1A in z-direction and locates at (-1 m, 1 m). The frequency of the source is 0.5 GHz. The surrounding regions are perfect match layer (PML) regions. When we place a conventional window in the free space, the light propagates through the window and reaches the lower space because the window is large enough (see in Fig. 3(b)). For comparison, if we replace the conventional window by the designed transformation media following the approach shown in Fig. 1, the light will be concentrated in the transformation media and also reach the lower space as if there is a big window opened for it (See in Fig. 3(c)). Here the continuity of the coordinate transformation in the positions ($y' = \pm \frac{d_1}{2}$) leads the device to be impedance-matching in the interfaces with the upper space and lower space. In contrast, if we take away the transformation media device, the information reaching the lower space will be greatly reduced (See in Fig. 3(d)).

In conclusion, we have proposed a kind of transformation media in a narrow slit, which can achieve the same transparency of a wide window. This kind of devices may find their applications in the industries and other fields. The simple case here is to turn a 2D narrow slit into a wider window. The same idea can be applied to a more complex three dimensional case, that is to turn a small hole into a bigger one.

**Acknowledgements**

This work was supported by the National Natural Science Foundation of China under grand No.10334020 and in part by the National Minister of Education Program for Changjiang Scholars and Innovative Research Team in University.

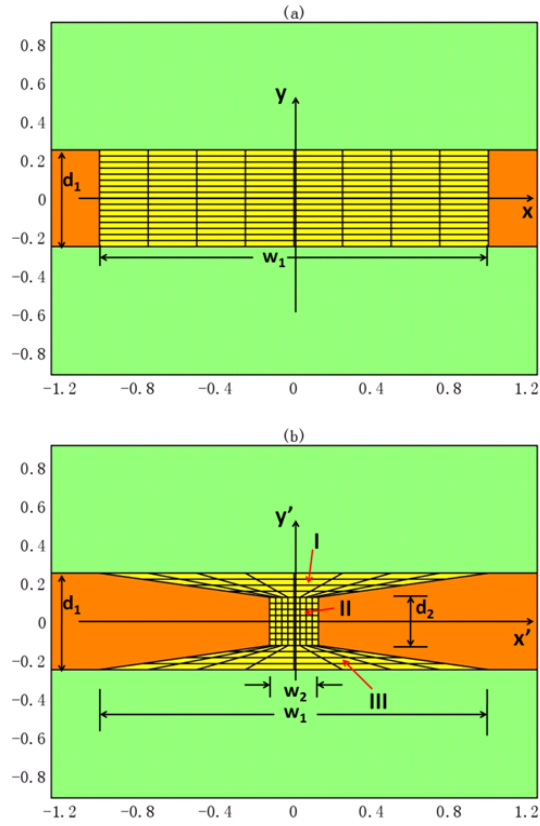

Fig. 1 (a) A rectangular free space sandwiched between two PECs, whose lengths of the sides are $w_1$ and $d_1$. (b) The rectangular free space is compressed into a bridge-shaped space. The "bridge" is another smaller rectangle whose lengths of the sides are $w_2$ and $d_2$. The original free space and the transformed space are highlighted by grids and yellow color. The green regions are vacuums. The orange regions are PECs.

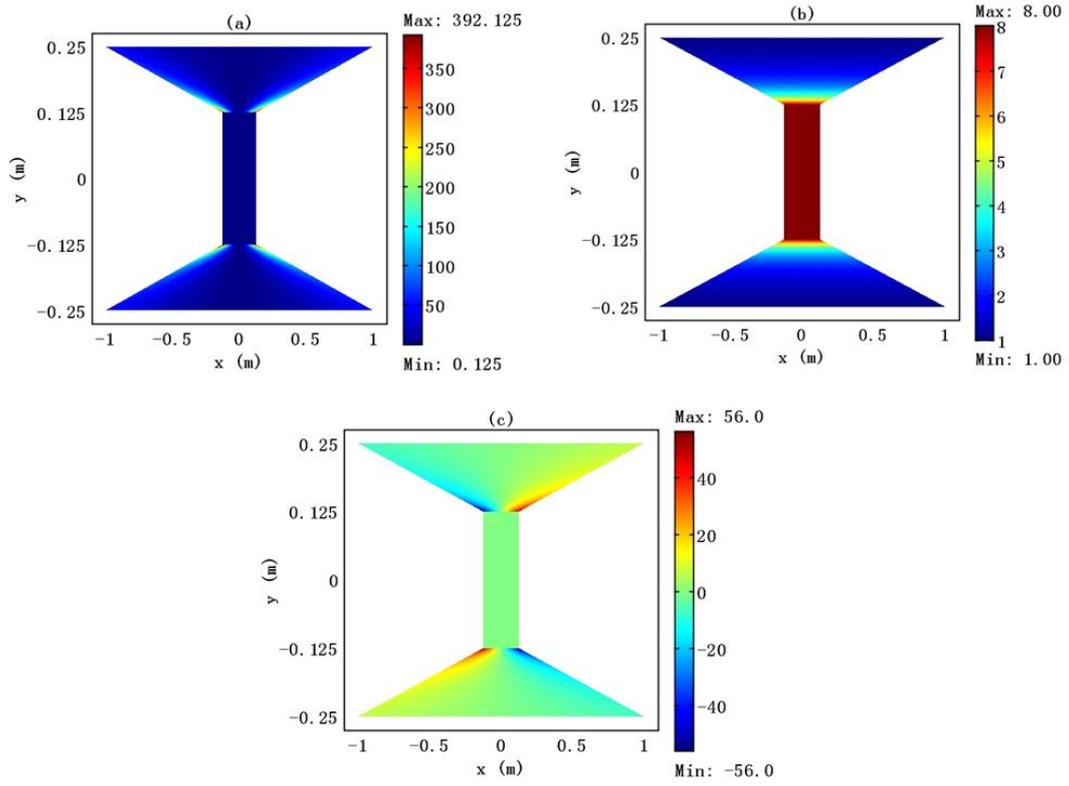

Fig. 2 Spatial dependence of the material parameters of the designed device for z-polarized TE wave, (a) $\mu_{xx}$, (b) $\mu_{yy} = \varepsilon_z$, (c) $\mu_{xy}$.

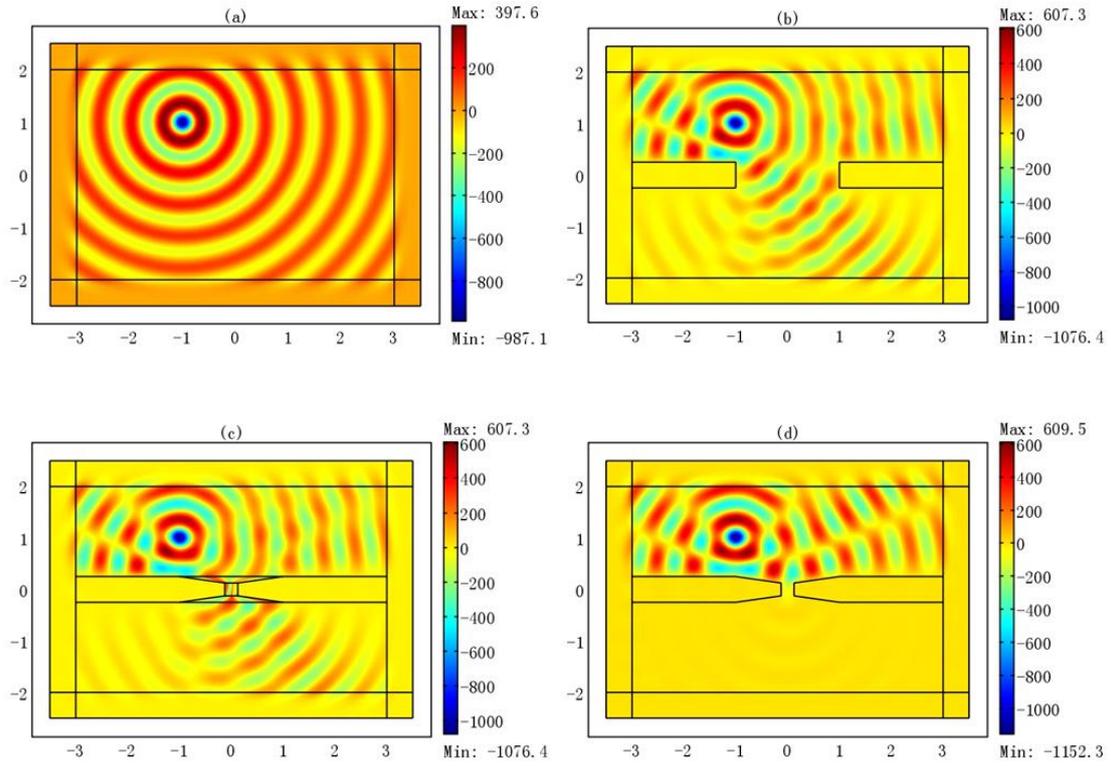

Fig. 3 (a) The electric field distribution for an electric line source in the free space. (b) The electric field distribution for an electric line source near a big "window". (c) The electric field distribution for an electric line source near the designed transformation media device. (d) The electric field distribution for an electric line source near the same small slit as in (c) without the transformation media in it.